\documentclass{SCIS2025}
\usepackage{enumitem}

\begin{document}

\ArticleType{POSITION PAPER}

\Year{2025}
\Month{}
\Vol{}
\No{}
\DOI{}
\ArtNo{}
\ReceiveDate{}
\ReviseDate{}
\AcceptDate{}
\OnlineDate{}
\AuthorMark{}
\AuthorCitation{}

\title{Towards Wireless Native Big AI Model: The Mission and Approach Differ From Large Language Model}{Towards Wireless Native Big AI Model: The Mission and Approach Differ From Large Language Model}

\author[1,2]{Zirui Chen}{}
\author[1,2,3]{Zhaoyang Zhang}{{ning\_ming@zju.edu.cn}}
\author[1,2]{Chenyu Liu}{}
\author[1,2,3]{Ziqing Xing}{}

\AuthorMark{Chen Z R}

\AuthorCitation{Chen Z R, Zhang Z Y, Liu C Y, et al}

\address[1]{College of Information Science and Electronic Engineering, Zhejiang University, Hangzhou {\rm310027}, China}
\address[2]{Zhejiang Provincial Laboratory of Multi-Modal Commun. Netw. and Intell. Inf. Process., Hangzhou {\rm310027}, China}
\address[3]{Institute of Fundamental and Transdisciplinary Research, Zhejiang University, Hangzhou 310058, China}

\abstract{
Research on leveraging big artificial intelligence model (BAIM) technology to drive the intelligent evolution of wireless networks is emerging. However, breakthroughs in generalization brought about by BAIM techniques mainly occur in natural language processing. There is a lack of a clear technical direction on how to efficiently apply BAIM techniques to wireless systems, which typically have many additional peculiarities. To this end, this paper reviews recent research on BAIM for wireless systems and assesses the current state of the field. It then analyzes and compares the differences between language intelligence and wireless intelligence on multiple levels, including scientific foundations, core usages, and technical details. It highlights the necessity and scientific significance of developing wireless native BAIM technologies, as well as specific issues that need to be considered for technical implementation. Finally, by synthesizing the evolutionary laws of language models with the particularities of wireless systems, this paper provides several instructive methodologies for developing wireless native BAIM.
}

\keywords{Wireless artificial intelligence, foundation model, wireless big AI model.}

\maketitle

\section{Introduction}
Given the outstanding potential of artificial intelligence (AI) technologies in complex feature extraction, high-dimensional data representation, and adaptive decision-making, the application of AI technology to resolving problems in wireless systems has been a critical topic in wireless technology research \cite{AI_for_5G, AI_for_B5G, AI_for_6G, AI_for_6G2, AI_for_6G3, AI_for_6G4}. However, simple migrating or reusing successful approaches from computer vision (CV) or natural language processing (NLP) to wireless systems tends to create serious bottlenecks in terms of performance and generalizability \cite{seq2seq}. This “incompatibility” is because AI for wireless is not directly equivalent to AI for NLP or CV. Indeed, it is necessary to incorporate electromagnetic physics and engineering methods, which makes modeling wireless AI substantially different from modeling human cognition and interaction capabilities in NLP and CV. High-quality wireless AI technologies must adapt to the characteristics of wireless data and the unique requirements of wireless systems, a concept known as becoming “wireless native.” {\color{black}In recent years, the incorporation of wireless native methods like physics-inspired structures \cite{seq2seq, siren, ode, mfcnet, cmixer, disentangled_representation} and cross-module combination \cite{channel_deduction, beam_and_feedback} has significantly improved the performance and generalization of wireless AI. Meanwhile, the introduction of prior information and advanced learning architectures can reduce the demand for training data and computational resources. These fundamental advantages are of great significance for advancing the usability of wireless AI, whether in centralized or distributed usage scenarios \cite{AI_for_B5G}.}

\begin{figure*}[htbp]
    \centering
    \includegraphics[width=0.94\linewidth]{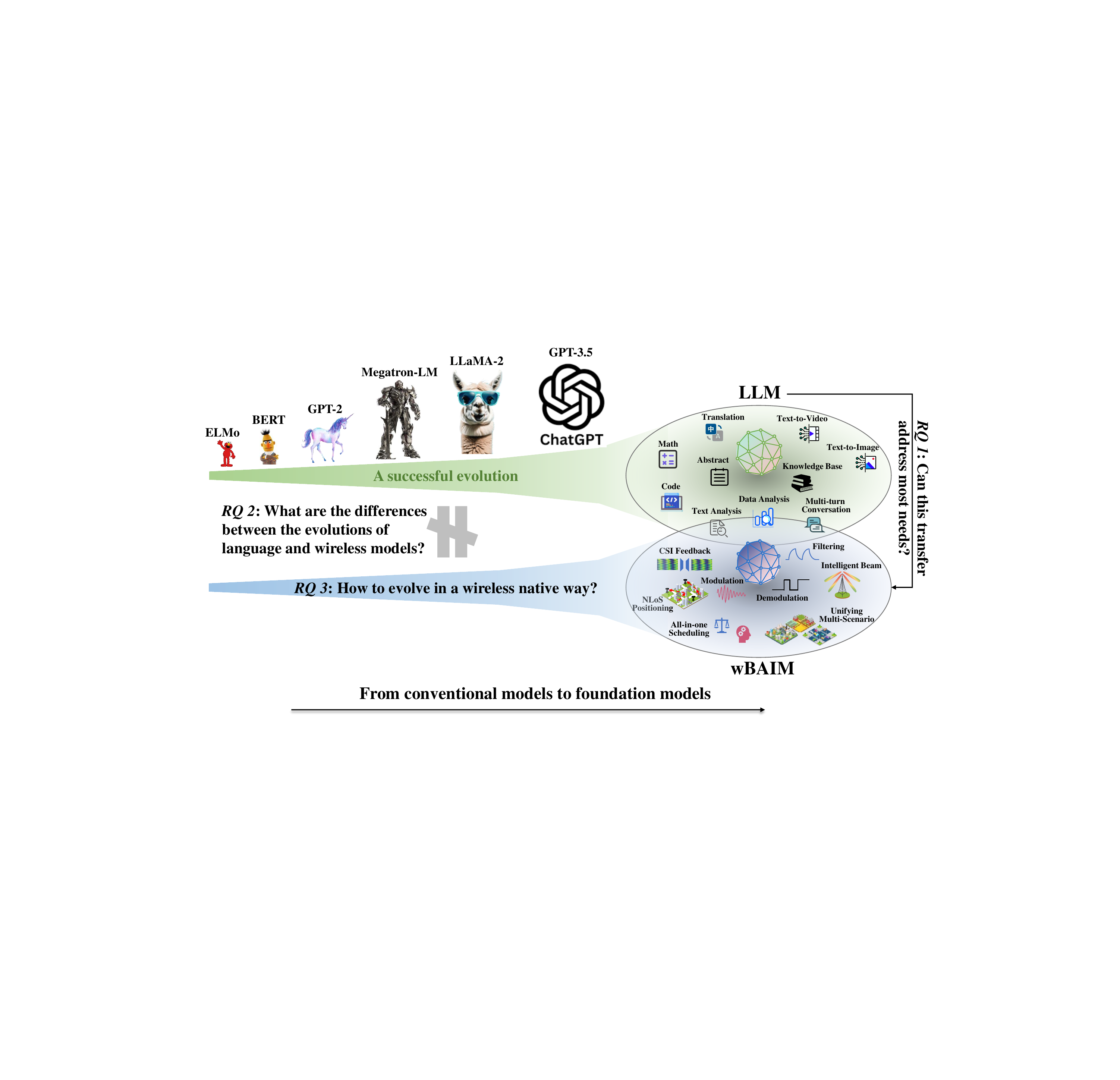}
    \caption{Research backgrounds and critical questions of `BAIM for wireless'.}
    \label{fig_question}
\end{figure*}

Although much progress has been made in wireless AI research, advancements have still only produced intelligence limited to specific tasks and scenarios, which creates obvious weaknesses in generalizability and stability. Meanwhile, in the NLP field, recent large language models (LLMs) have demonstrated the outstanding potential of big AI models (BAIMs) in broad generalization and high reliability. With the help of a pre-trained LLM, numerous language tasks like translation, question answering (Q\&A), and summarization can be accomplished with state-of-the-art performance \cite{FoM}. Moreover, the emergence of advanced LLMs like DeepSeek has significantly reduced the training and deployment costs of such large-scale AI models \cite{DeepSeekV2,DeepSeekV3,Qwen}, further expanding the applicability of these cross-task general intelligence.
Wireless AI desperately needs similar advancements in generalization. Researchers in \cite{R28} first defined wireless BAIM (wBAIM) as having the typical features of integrating multitasks, unifying multiscenarios, and all-in-one scheduling. This definition emphasizes the development of a BAIM that focuses on the physical nature of wireless data and the fundamental requirements of wireless systems.
Realizing these promising visions would represent a quantum leap in functionality, universality, and reliability for wireless AI, but it requires the support of comprehensive, in-depth technical exploration. Although \cite{R28} has summarized the technical challenges and potential solutions of wBAIM, several issues almost immediately emerged with wBAIM that still require clarification and supplementation. These issues informed the research questions (RQs) for this study (see Fig. \ref{fig_question}):
\begin{enumerate}[label=\normalsize$\bullet$]
\item \emph{RQ 1}: What are the most necessary and significant aspects when investigating wBAIM in the context of powerful LLMs and their transferability?
\item \emph{RQ 2}: What are the technical peculiarities of wBAIM, especially in comparison to conventional wireless models and existing LLMs?
\item \emph{RQ 3}: What methodologies for developing wBAIM can be derived by synthesizing the evolutionary laws of language models and the peculiarities revealed by \textit{RQ 2}?
 \end{enumerate}

To answer these RQs, this paper is organized as follows. Section \ref{section2} provides an overview of the current research landscape and recent advancements in the field. Next, in Section \ref{section3}, the fundamental differences between language and wireless are analyzed, highlighting the necessity, challenges, and value of studying wBAIM. Section \ref{section4} then examines, from a system-level perspective, the unique technical considerations essential for wBAIM research. Section \ref{section5} presents several forward-looking insights into how to develop wBAIM. Finally, the conclusions are drawn in Section \ref{section6}.

\section{Literature Review}\label{section2}
\begin{figure*}[htbp]
    \centering
    \begin{minipage}[b]{0.28\linewidth}
        \centering
        \includegraphics[width=\linewidth]{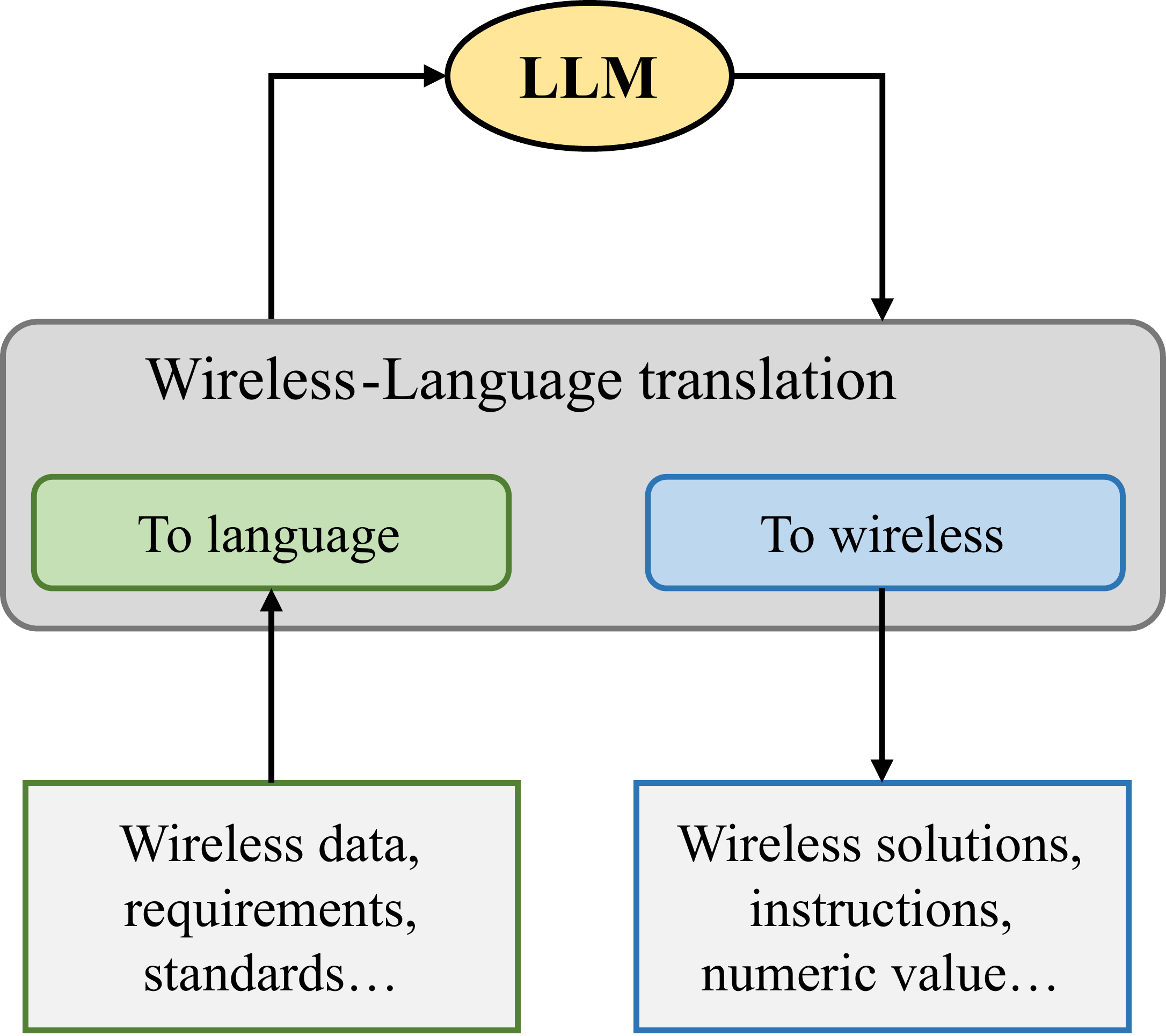}
        \caption*{(a) Wireless LLM.}
    \end{minipage}%
    \begin{minipage}[b]{0.04\linewidth}
        \centering ~~
    \end{minipage}%
    \begin{minipage}[b]{0.28\linewidth}
        \centering
        \includegraphics[width=\linewidth]{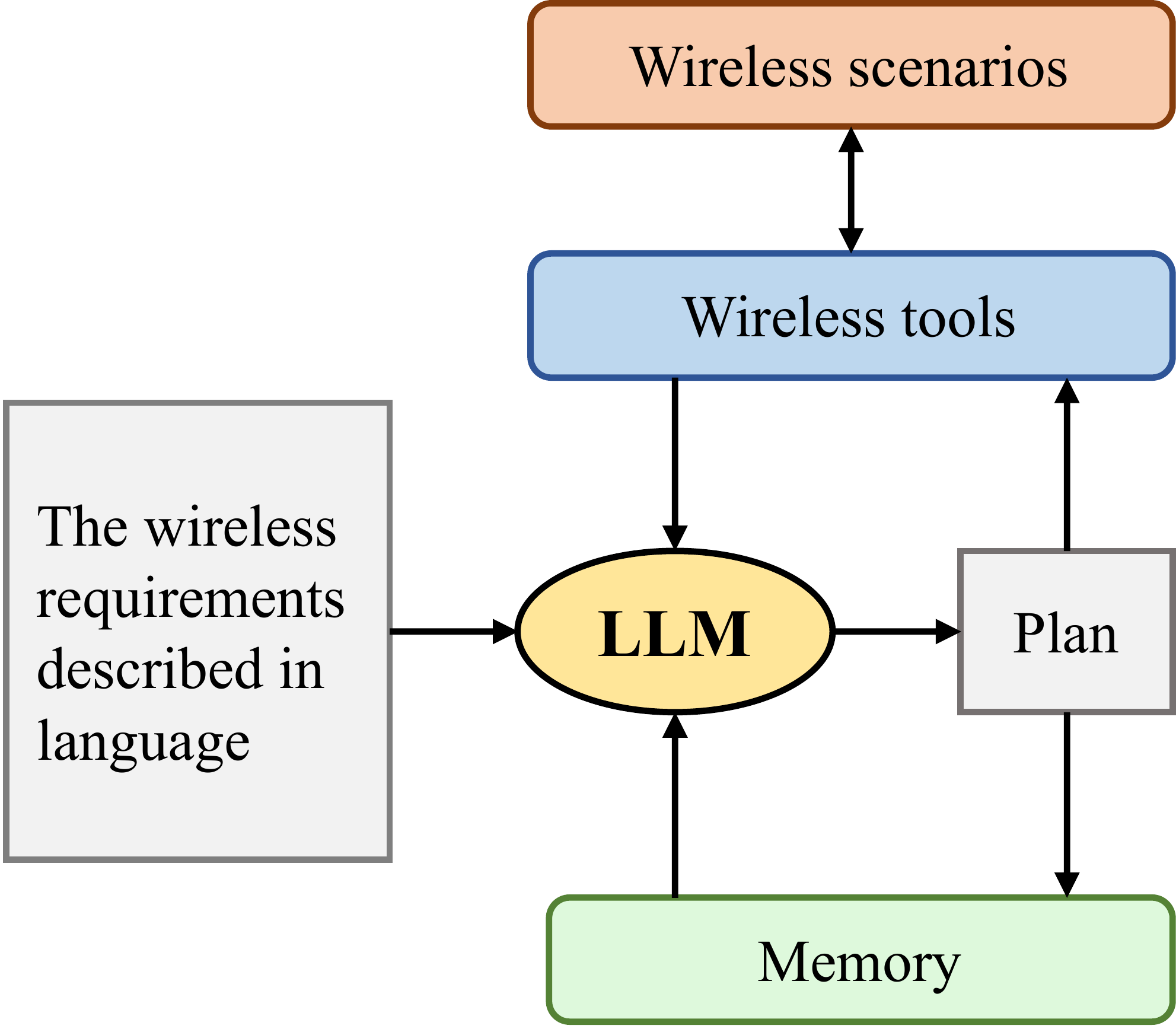}
        \caption*{(b) LLM-based wireless agent.}
    \end{minipage}%
    \begin{minipage}[b]{0.04\linewidth}
        \centering ~~
    \end{minipage}%
    \begin{minipage}[b]{0.34\linewidth}
        \centering
        \includegraphics[width=\linewidth]{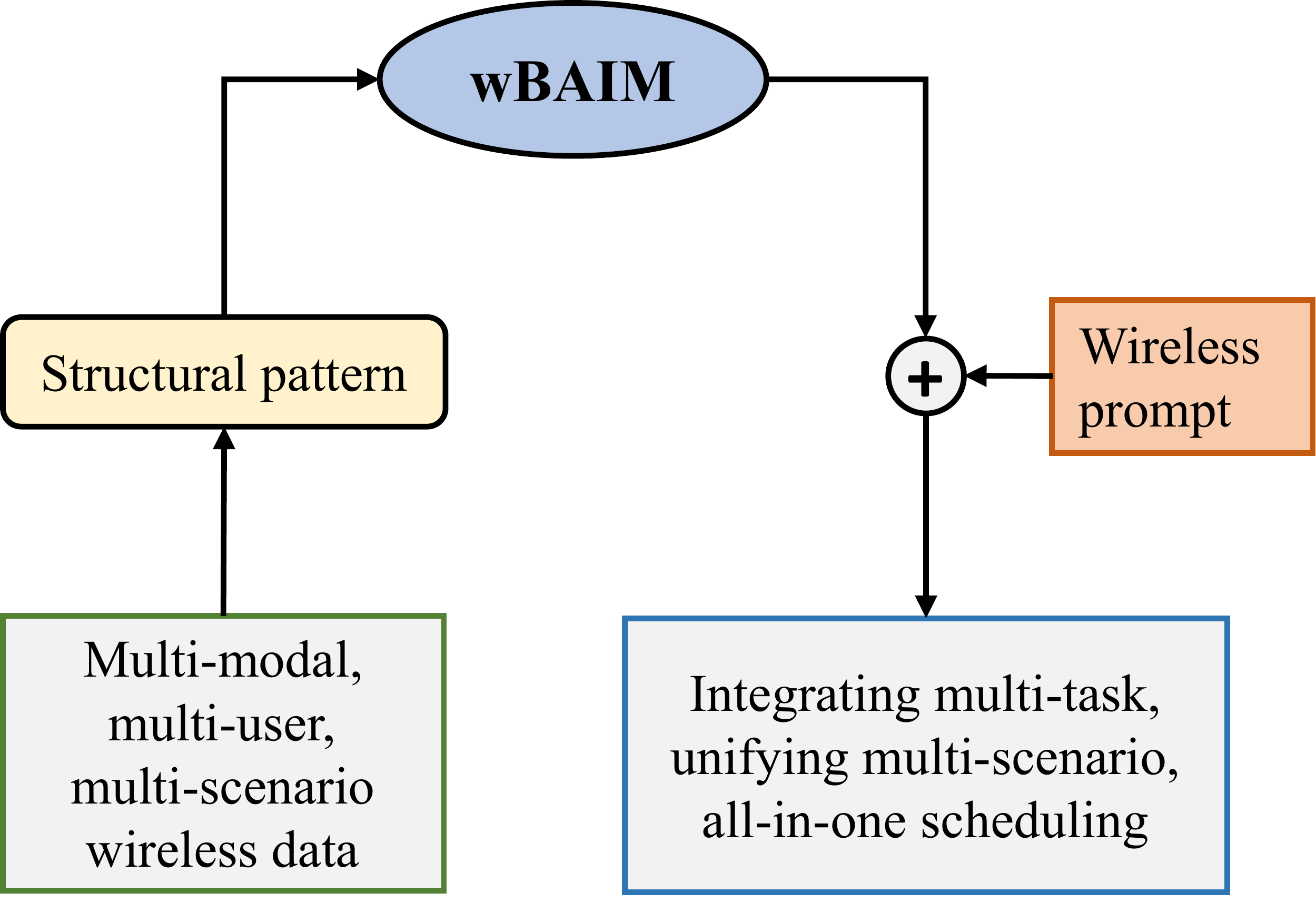}
        \caption*{(c) wBAIM.}
    \end{minipage}
    \caption{The common research paradigms on `BAIM for wireless'.}
    \label{fig_categorization}
\end{figure*}

The success of LLMs has motivated numerous researches on BAIM for wireless to accelerate wireless systems towards ubiquitous intelligence. Existing researches can be generally categorized into the following three research paradigms: wireless LLM, LLM-based wireless agent, and wBAIM. The overall architectures of these three paradigms are shown in Fig. \ref{fig_categorization}.

\subsection{Wireless LLM}\label{section2.1}
Wireless LLM refers to applying existing LLMs to wireless problems through reuse, fine-tuning, retrieval augmented generation (RAG) techniques, and so on. These methods translate wireless system requirements into language-based descriptions, positioning them within (or approximately within) LLMs’ generalization capabilities. Leveraging LLMs’ advanced information comprehension and vast knowledge base, initial successes have been achieved in wireless systems with protocol understanding, command interaction, and resource allocation \cite{R18, R22}. Moreover, some researchers have applied LLMs to NLP-related tasks in wireless systems, such as text categorization \cite{R5}, text summarization, and Q\&A \cite{R6, R7}, with specific applications including the creation of AI platforms for customer interaction, virtual assistants for technicians, network diagnostic management, and telecom-standard Q\&A bots \cite{R8, R9, R10}. While these studies have demonstrated the value of LLMs in wireless systems, they primarily remain within the scope of traditional NLP. Furthermore, LLMs have been applied to practical requirements in wireless systems such as power control, intrusion detection, resource allocation, and symbol demodulation \cite{R23, R24, R25, R26}. By constructing appropriate prompts, LLMs based on in-context learning can understand wireless problems to a certain extent, thus enabling initial wireless-specific functions without the need for parameter adjustments.

Meanwhile, to further enhance the applicability of LLMs in wireless communications, some researchers have begun exploring the integration of wireless domain knowledge into LLMs while focusing on the role of multimodal information \cite{R11}. These researchers have first pre-trained generative models on specific telecom knowledge datasets, followed by supervised finetuning (SFT) of the models with telecom instruction sets \cite{R12}, and aligning the model’s outputs with human preferences through reinforcement learning from human feedback (RLHF) \cite{R13, R14, R15}. In addition to direct retraining, the RAG mechanism and the LangChain framework have also been identified as important supplements for a telecom knowledge base \cite{R16, R17, R19, R20, R21}. Additionally, some researchers have constructed foundational telecom text datasets to serve as material for expanding LLMs’ wireless knowledge. These include SPEC5G, a text categorization dataset derived from 3rd Generation Partnership Project (3GPP) technical reports \cite{R1}; TeleQnA, a Q\&A dataset extracted from technical documents and research materials \cite{R2}; TSpec-LLM, a dataset based on 3GPP specifications that preserves table formulas \cite{R3}; Tele-Data, a LaTeX-formatted dataset obtained from sources like arXiv, 3GPP standards, and Wikipedia; and Tele-Eval, a Q\&A evaluation dataset \cite{R4}.

\subsection{LLM-Based Wireless Agent}\label{section2.2}

From the perspective of model architecture, LLMs are primarily designed to meet the requirements of language tasks. As a result, they have inherent limitations in data representation, inference patterns, and other aspects when applied to wireless systems. To address these shortcomings, some researchers have begun integrating AI agent technology with wireless LLMs, exploring the concept of LLM-based wireless agents \cite{R32}. Unlike traditional wireless LLM, an LLM-based wireless agent uses LLM as a core component, combined with various functional modules to form a complete wireless agent. In \cite{R33}, the authors proposed a 6G LLM Agent, which utilizes an LLM pre-trained on a wireless text dataset as its "brain" for environment perception and action planning, and retrieves available tool libraries to accomplish wireless tasks. \cite{R35} further introduced the experience accumulation module to enable the continuous evolution of the AI agent. Additionally, in \cite{R34, R36},  researchers further analyze the significance of multi-agent collaboration in wireless systems.
{\color{black} It is important to note that, by invoking external tools, LLM-based agents can potentially outperform wireless LLMs in terms of functionality. However, the upper limits of their understanding and decision-making capabilities are still constrained by the LLM itself. As such, they still struggle to serve complete  wireless systems or solve wireless problems that cannot be solved by conventional algorithms. Moreover, the intricate logical architecture of the agent further increases the reasoning complexity beyond that of the LLM alone. }

\subsection{wBAIM}\label{section2.3}
Although LLM-based methods have been somewhat explored, the current technical progress primarily focuses on instruction design, resource scheduling, and a few simple wireless problems. {\color{black} These approaches still lack effective solutions for numerous tough wireless challenges, such as scenario sensing and high-rate transmission. This limitation arises from two primary factors: firstly, linguistic expressions are frequently insufficient in precisely characterizing multimodal signals and their intricate correlations; secondly, the language intelligence derived from textual training lacks adequate knowledge reserves to comprehend the data structures and underlying mechanisms inherent in wireless data.} Therefore, instead of embracing LLMs, some researchers have explored BAIM in a wireless native way.
Wireless native AI refers to an intelligent framework based on regular patterns of AI technologies, such as deep neural networks, while incorporating the peculiarities of wireless systems. It tailors model design to the nature of wireless data and system requirements, and adapts through re-learning on wireless data. Conventional wireless AI has also followed this approach in many studies. For example, in model design, researchers initially leveraged the numerical characteristics of wireless data, such as exploiting channel sparsity in the angle-delay domain and transforming it into the angle domain to make it more image-like \cite{csinet, cnn_positioning}. Later, researchers delved deeper into the physical mechanisms underlying the numerical characteristics and employed more fundamental, physics-inspired techniques to guide model optimization \cite{seq2seq, siren, ode, mfcnet, disentangled_representation}. In terms of architecture, some cross-module \cite{channel_deduction, beam_and_feedback} and cross-layer \cite{Federated_URA} joint techniques have also been developed to facilitate applications in wireless systems.

However, given that wBAIM has a broader functional scope, relying solely on existing wireless AI technologies is insufficient. Therefore, in \cite{R28}, the authors first outlined three key generalizability indicators of wBAIM: multi-task integration, multi-scenario unification, and all-in-one scheduling. Meanwhile, \cite{R28} also provides a demonstrative case study where a CMixer \cite{cmixer} model is trained across multiple scenarios to simultaneously accomplish high-dimensional channel estimation and channel compression feedback tasks, illustrating the feasibility and significance of wBAIM.
Building on this basis, \cite{R29} introduced SWTCAN, a network based on the advanced swin transformer, which integrates downlink pilot patterns, channel estimation, and uplink channel feedback into a joint design. \cite{R30} proposed Csi-LLM, a downlink channel prediction model based on GPT2, which supports variable-length inputs and enables continuous prediction. Similarly, in \cite{R31}, a pre-training and fine-tuning based channel prediction method was presented, enhancing performance in few-shot learning. Meanwhile, \cite{R27} evaluated the similarity and diversity across multi-scenario datasets, offering valuable insights into developing multi-scenario generalization techniques from a data-based perspective.
\vspace{0.5em}
\begin{figure*}[htbp]
    \centering
    \begin{minipage}[b]{0.55\linewidth}
        \centering
        \includegraphics[width=\linewidth]{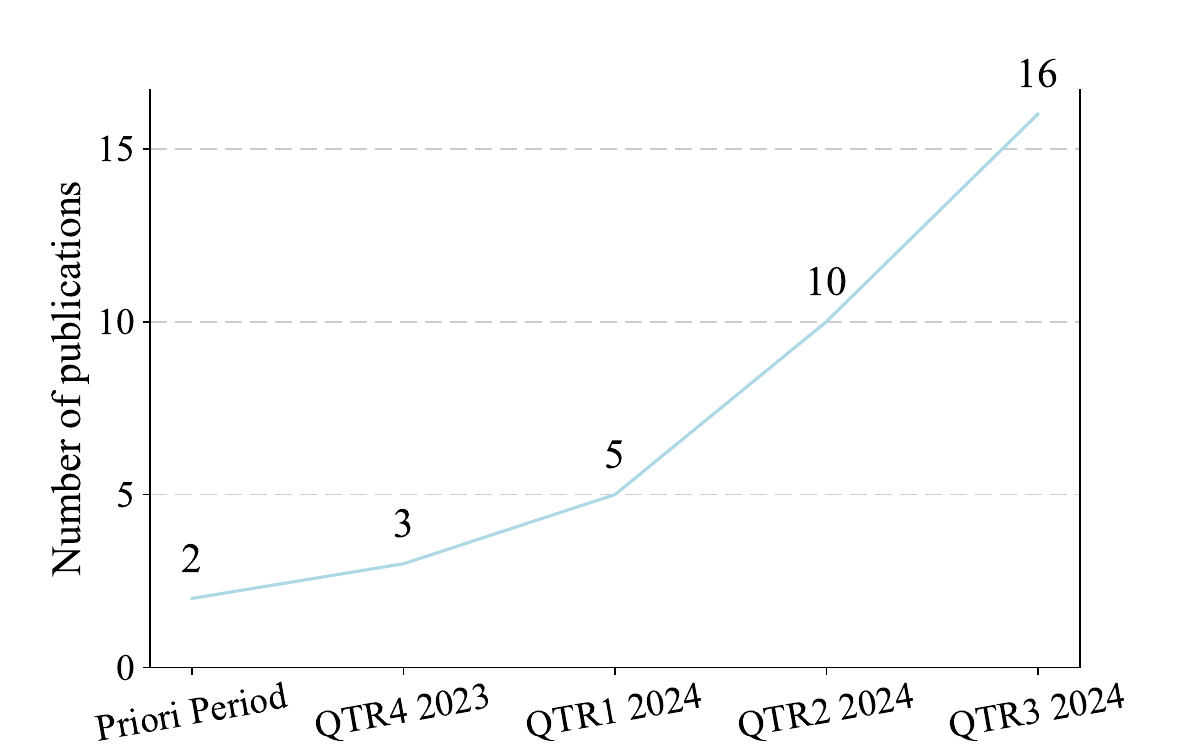}
        \caption*{(a) Recent quantitative trends.}
    \end{minipage}%
    \begin{minipage}[b]{0.04\linewidth}
        \centering
        ~
    \end{minipage}%
    \begin{minipage}[b]{0.4\linewidth}
        \centering
        \includegraphics[width=\linewidth]{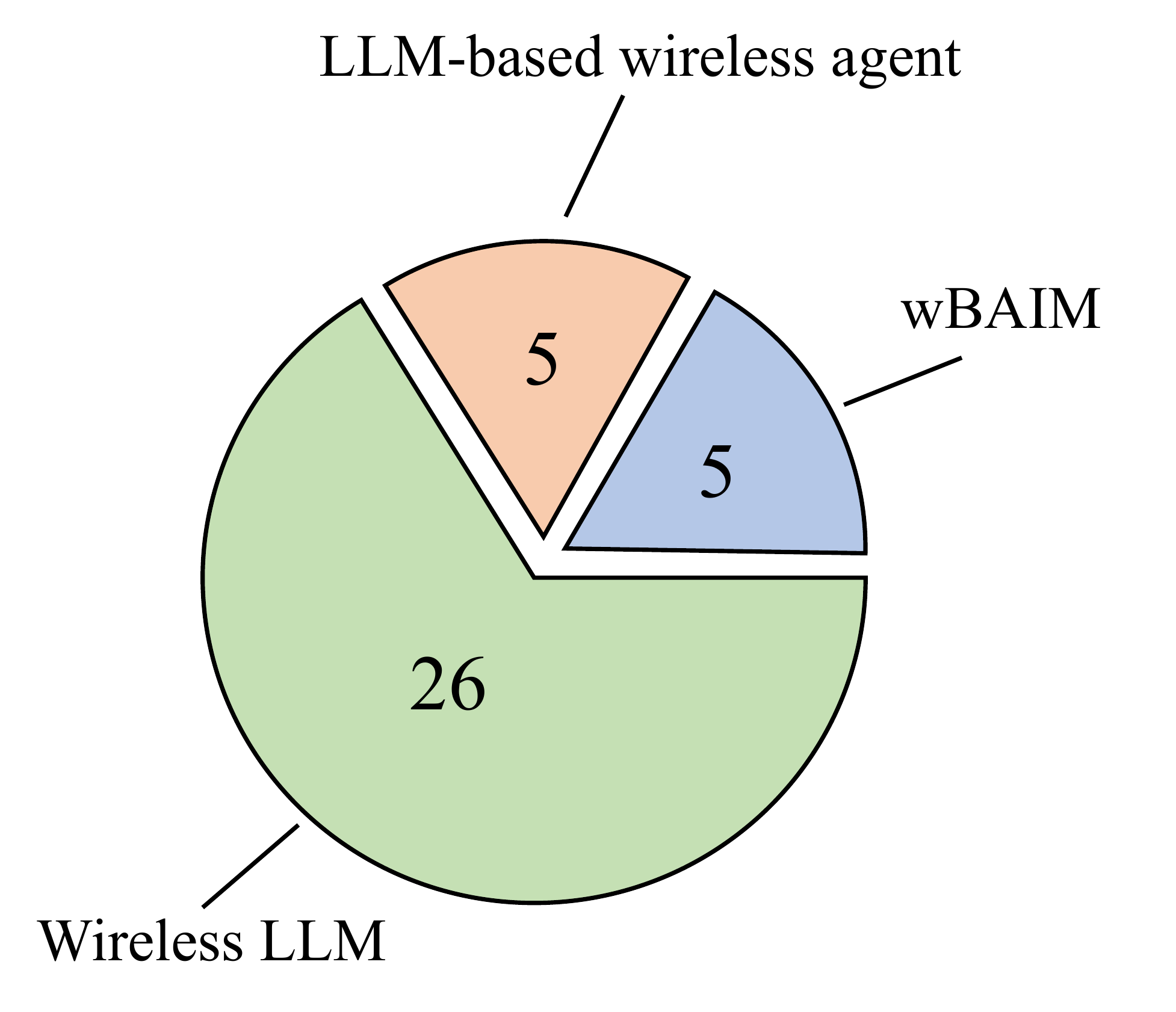}
        \caption*{(b) Categorization of studies.}
    \end{minipage}
    \vspace{0.5em}
    \caption{Statistics of reviewed publications related to `BAIM for wireless'.}
    \label{fig_stactisitics}
\end{figure*}
\subsection{Summary of the Literature}\label{section2.4}
Fig. \ref{fig_stactisitics} presents statistical information on the 36 publications related to BAIM in wireless systems included in this paper. It shows that research interest in BAIM for wireless is growing, and there is a growing consensus on the importance of BAIM techniques for the evolution of 6G. However, studies based on LLMs dominate among the existing paradigms, while research on wBAIM remains limited, and technical advances in wBAIM are still in the early stages. Furthermore, current LLM-based solutions primarily focus on a narrow set of wireless tasks, due to substantial differences in core principles and application contexts between language and wireless domains. The integration of BAIM into wireless systems has not yet reached the depth and breadth it has achieved in language and other fields. Therefore, the following sections highlight the necessity and significance of wBAIM, offering critical insights into its unique challenges and methodologies to foster a deeper integration of BAIM and wireless technologies.

\section{Comparisons with LLM: The Ambitions of wBAIM}\label{section3}
\subsection{Differences in Intelligence Orientation}\label{section3.1}
Leveraging existing LLMs, the research value of integrating BAIM techniques and wireless networks can be initially evaluated with low resources and uncomplicated technology adaptation, which is an important motivation for why existing research primarily investigates wireless LLM and LLM-based wireless agent. However, there are many intuitionistic differences between language and wireless. What is the scientific nature of such differences? Can one expect to build complete universal wireless intelligence based on LLM's transfer in the presence of such differences? This is one of the important questions to be answered in this paper.

\begin{figure*}[htbp]
    \centering
    \includegraphics[width=0.98\linewidth]{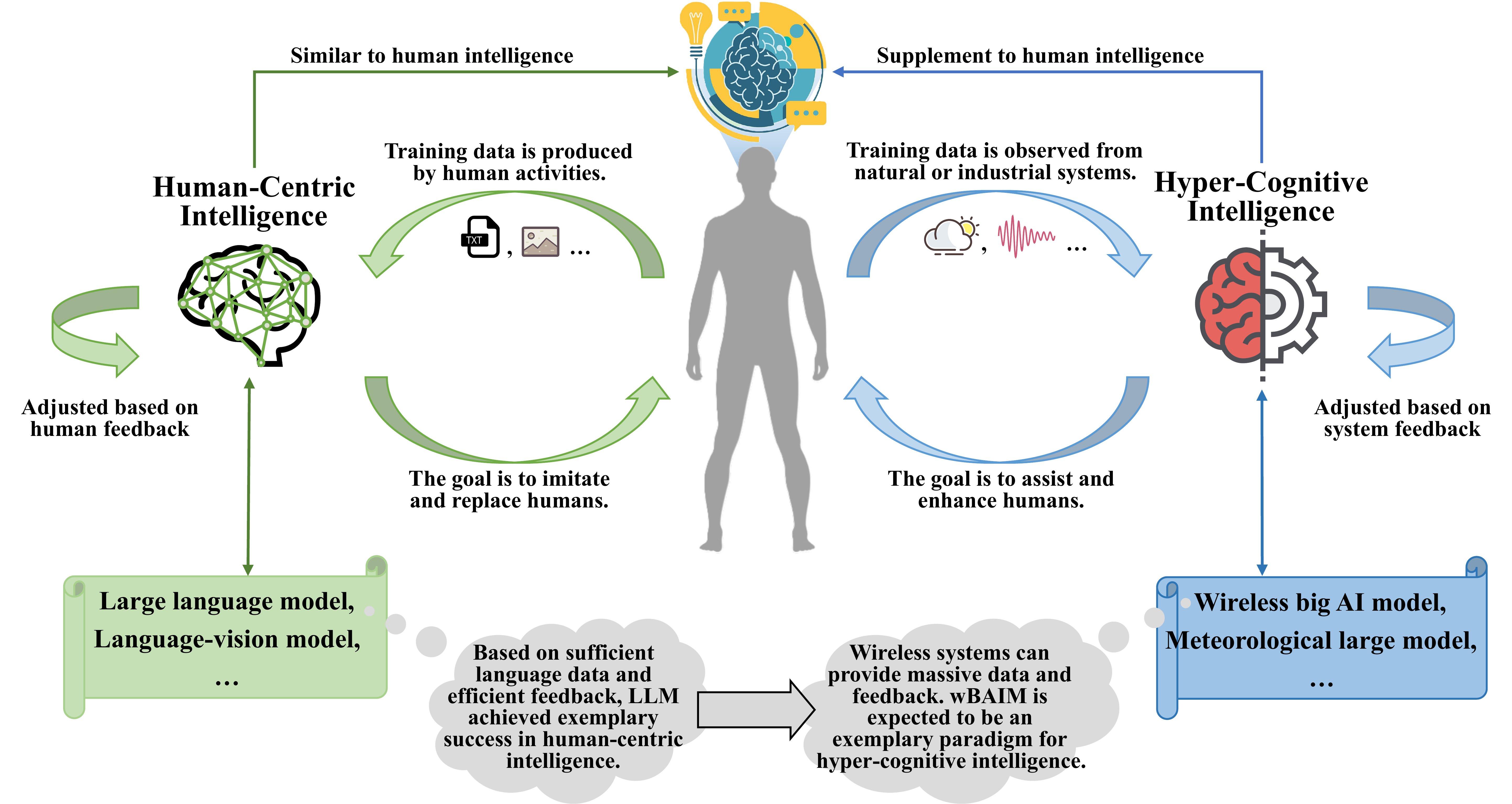}
    \vspace{0.3em}
    \caption{A graphical illustration of human-centric and hyper-cognitive intelligence, which reflects substantial differences in intelligence orientation between LLM and wBAIM.}
    \label{fig_comparsions}
\end{figure*}

In fact, the intelligence orientation and applications of AI technologies can be divided into two categories based on human cognition, which we term “human-centric” and “hyper-cognitive,” as illustrated in Fig. \ref{fig_comparsions}. The human-centric model primarily learns from human-generated data, such as images and language, and is continually refined through human feedback, such as reinforcement learning from human feedback (RLHF), ultimately aiming to replicate humanoid intelligence. In this sense, the "teacher" of the human-centric model is humanity itself, and its ultimate goal is to replace and liberate humans. However, due to the inherent limitations of biological intelligent agents in complex perception and intensive computation, humans lack the deep understanding and control over many natural and industrial systems. As a result, hyper-cognitive intelligence becomes essential. The hyper-cognitive model primarily learns from phenomena in natural and industrial systems, such as weather patterns and electromagnetic signals, and is optimized through system feedback, ultimately producing intelligence that surpasses the limits of human cognition. In this process, humans typically act as assistants, observing and interacting with information, while the model draws knowledge from systematic phenomena to help humans understand and manipulate complex scientific laws beyond their existing cognitive frameworks.

The distinction between these two forms of intelligence can be intuitively grasped by referencing human intelligence. Rooted in human-activity-and-feedback-based learning, this type of intelligence is inherently constrained by the limits of human cognition, making it challenging to fully address problems that exceed human understanding. Intuitively, language is a human-centric domain, whereas electromagnetic sciences and systems, which are not directly perceivable or recognizable by humans, represent hyper-cognitive challenges. Consequently, while LLMs excel at language-related tasks, they struggle to fully engage with electromagnetic intelligence—an area even beyond human capability. The identified differences address Question 1: LLMs and their transfer can simplify human interaction with wireless systems and manage certain simple wireless use cases. However, for complex wireless problems that are difficult to solve even with the mathematical and physical methods available to humans, LLMs cannot provide a complete solution.
\subsection{Scientific Significance of wBAIM}\label{section3.2}
Given these substantial differences, an in-depth understanding of electromagnetic laws is essential to more freely and accurately process, represent, and utilize electromagnetic signals. This necessitates the development of a BAIM that is native to electromagnetic sciences and systems—referred to as wBAIM. However, the black-box nature of AI models, coupled with the limitations of human cognition, makes it impossible to directly assess whether a given model has truly acquired the required capabilities. As a result, implicit assessment metrics become essential. This is precisely one of the reasons why \cite{R28} defines multi-task integration, multi-scenario unification, and all-in-one scheduling as the key features of wBAIM. {\color{black} If a single model can be used to generalize across multiple wireless tasks and scenarios, it must inherently capture the commonalities among them—that is, the general electromagnetic laws. In other words, this is the “compression is intelligence” in the wireless context. Likely, this intelligence not only serves wireless communication systems but also provides technical supports for other electromagnetics-related fields such as remote sensing \cite{remote_sensing,meta_sensing} and radar systems \cite{radar_system}.

Compared to human-centric intelligence, hyper-cognitive intelligence is also crucial for the advancement of human society, though its development is still in relatively early stages. Given that wireless network is a widely-used, scientifically well-defined, and highly digitized field within common natural and industrial systems, their advantages in data, computational power, and system feedback are particularly pronounced. As such, wBAIM serves as an ideal candidate for studying hyper-cognitive intelligence, with significant potential to lead technological breakthroughs and provide a valuable reference framework for AI-driven science and engineering.}

\section{Peculiarities of wBAIM}\label{section4}

\begin{figure*}[htbp]
    \centering
    \includegraphics[width=0.92\linewidth]{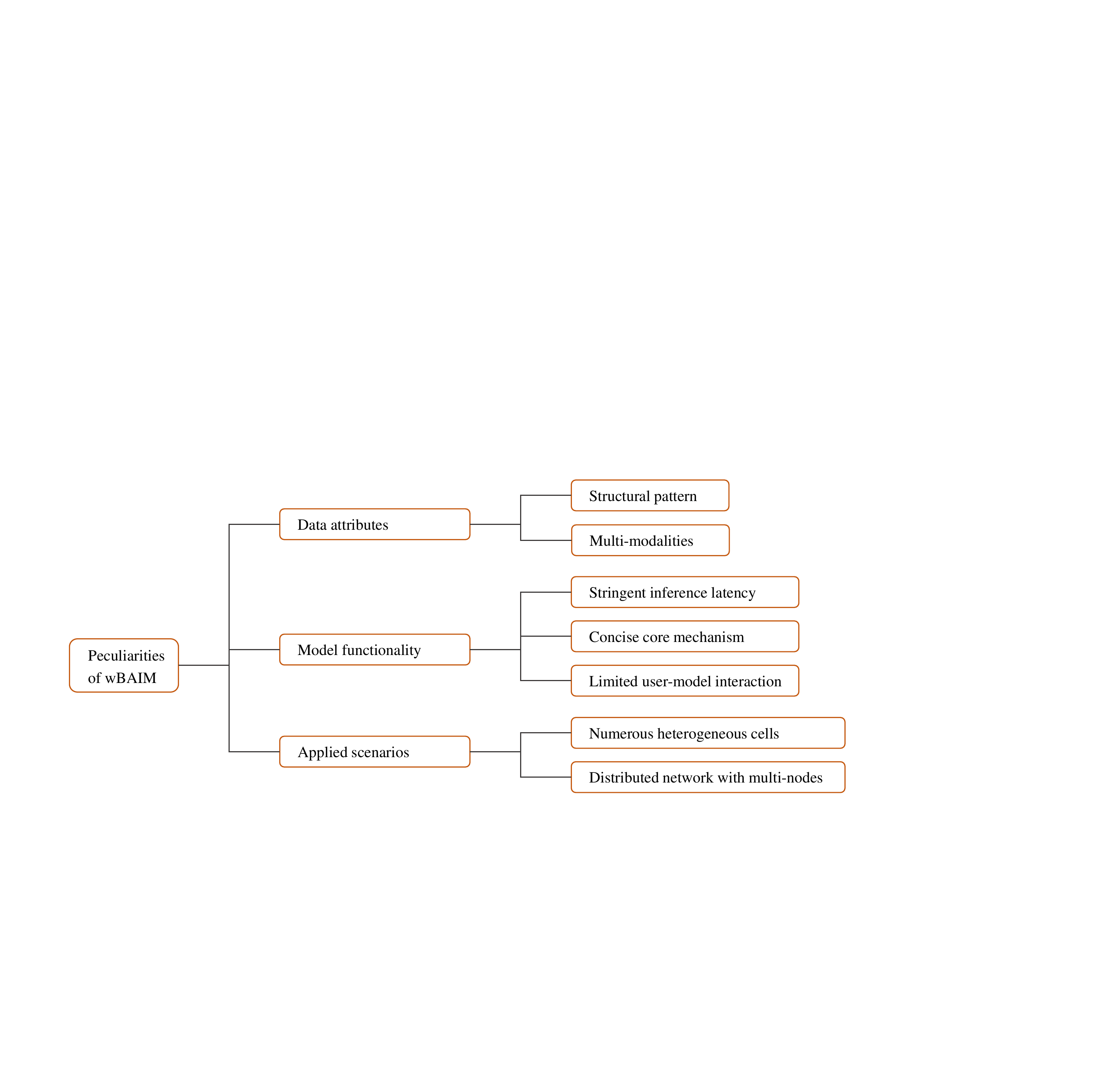}
    \caption{The technical peculiarities of wBAIM in terms of data, model, and application.}
    \label{fig_peculiarities}
\end{figure*}

As discussed in Section \ref{section3}, the intelligence inherent in natural language does not entirely satisfy the functional requirements of wireless systems. To fully leverage the potential of BAIM in wireless systems, it is necessary to develop wBAIM in a wireless native manner. Before creating practical implementation methods, we must evaluate critical aspects of AI technology development, including data, models, and applications, as summarized in Fig. \ref{fig_peculiarities}. This evaluation will help us to understand how wBAIM differs from conventional wireless models and LLMs.

\subsection{Data Attribute}\label{section4.1}
In conventional wireless models, the focus is often limited to specific tasks, meaning these models typically process only unimodal data in input or output. However, one key feature of wBAIM is its integration of multiple wireless tasks, which requires representing and processing multimodal data, such as CSI, modulation symbols, and positional coordinates. In contrast, language data are inherently unimodal, and multimodal data processing is not a crucial requirement in LLMs. Other BAIMs, such as language-vision models (LVMs), primarily rely on the alignment of semantic information to handle multimodal data, yet there remains a need to better exploit the correlations and complementarities among these modalities. As a result, addressing the multimodality of wireless data requires further technical research. Moreover, wireless data are typically raw data obtained through sampling and observation, which impose strict logical order, correspondence, and structural properties. In comparison, language data, which have been pre-processed by humans, do not require the same strict order and logic, as they can convey semantic information accurately even when somewhat rearranged. This fundamental difference places higher demands on feature extraction and representation learning structures in wBAIM.

\subsection{Model Functionality}\label{section4.2}

Unlike language models, which encompass a wide variety of knowledge, the fundamental laws of electromagnetic waves are relatively simple. The complexity of wireless signals often comes from the deep coupling of these simple laws with scattering environments and transmission mechanisms. This characteristic should be reflected in the development of wBAIM, i.e., the key to the functionality of the model is its reasoning ability rather than its memorization. Specifically, this involves how to guide the model to understand the concise mechanism behind complex data and how to couple the learned mechanism with the features of the target scenario.

Additionally, wireless systems require real-time transmission, making it essential to strictly control inference time to avoid issues like outdated state information and additional transmission latency. This means that wBAIM urgently needs lightweight reasoning mechanisms under a large information capacity. Furthermore, since obtaining real-time state information in wireless systems often incurs additional signaling overhead, the interactions between wBAIM, the system, and users are inherently limited. As a result, a mechanism capable of deep information extraction and high-density data representation is more needed for wBAIM than for processing extremely long information sequences.

\subsection{Applied Scenarios}\label{section4.3}

The unique networking architecture of wireless systems also brings peculiarities to wBAIM. {\color{black} On the one hand, since the major objective of wBAIM is to unify various scenarios and provide all-in-one scheduling, it requires wBAIM to be generalizable across diverse and heterogeneous cellular scenarios. The transmission configurations and scatterer environments within different scenarios are usually different, making it a key technical challenge to enable wBAIM to quickly adapt to various scenarios. On the other hand, as wireless networks are evolving towards decentralization, the wireless data and computational resources are inherently distributed across scenarios. This decentralized nature makes it impractical for wBAIM to perform training and inference in a fully centralized manner, which differs from the mainstream LLM paradigm.} Therefore, developing an efficient, multi-node architecture for collaborative computation is another critical technical requirement for advancing wBAIM.

\section{Methodologies to Build wBAIM}\label{section5}

\begin{figure*}[htbp]
    \centering
    \includegraphics[width=0.92\linewidth]{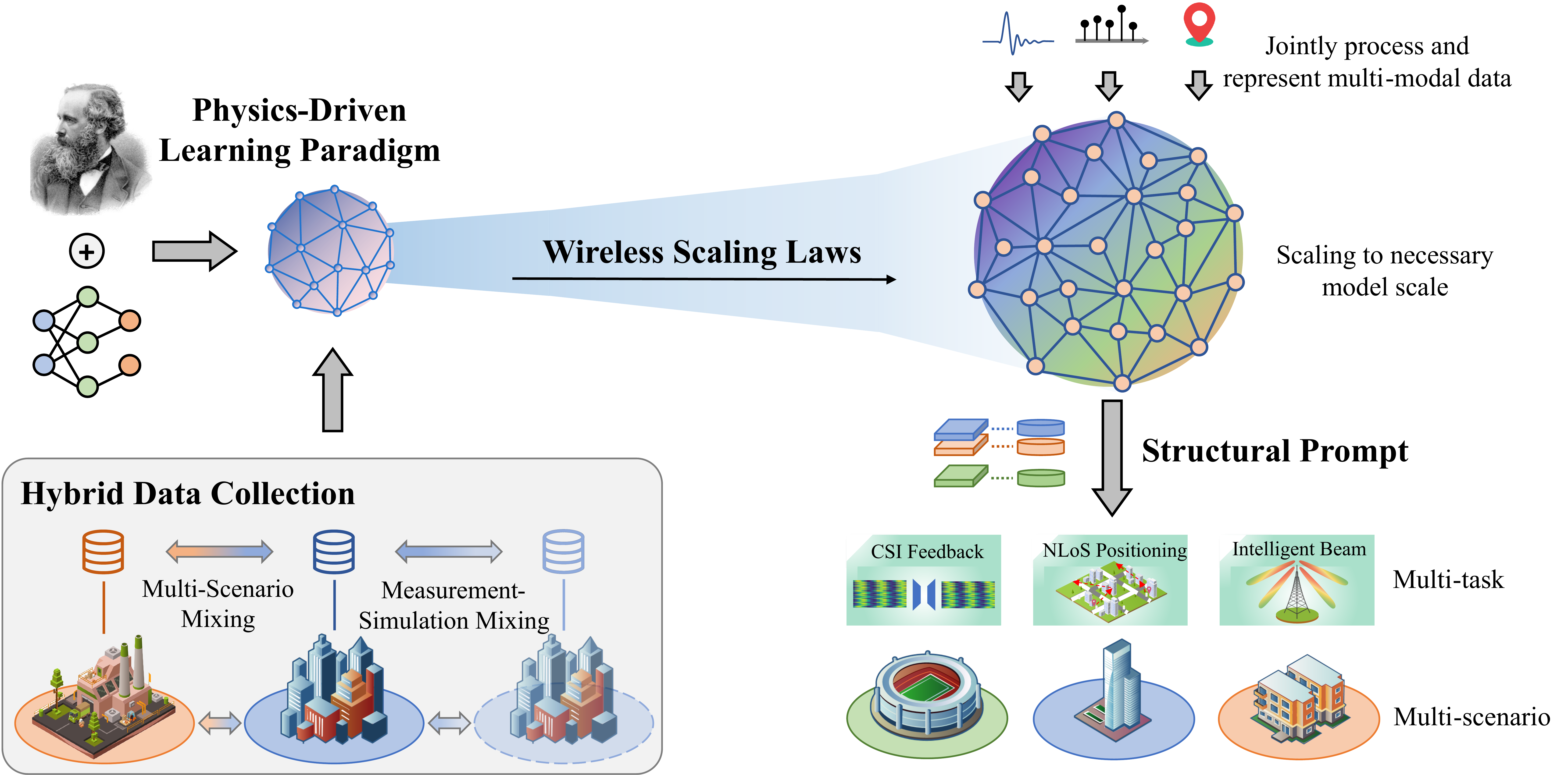}
    \vspace{0.3em}
    \caption{Systematic methodologies in developing wBAIM.}
    \label{fig_methodologies}
\end{figure*}

Despite the essential differences between wireless and language, scientific development often has strong commonalities across different domains. The successful evolution of LLMs from conventional language models is essentially driven by four key elements: the extensive data as a source of intelligence, the learning paradigm oriented towards general intelligence, the emergence effect brought about by the expansion of model structure and scale, and the efficient method for use case adaptation. In this paper, we map these four elements to the wireless case, merge the evolutionary laws of LLMs with wireless peculiarities, and provide a methodology to guide the development of wBAIM, as shown in Fig. \ref{fig_methodologies}.

\subsection{Hybrid Data Collection}\label{section5.1}

A rich knowledge base of training data is the foundation for BAIM to achieve wide-ranging intelligence. However, since wBAIM aims to serve massive heterogeneous cellular scenarios, the wireless datasets used for pre-training must adequately reflect the scenario diversity, which requires mixing the data from multiple scenarios. Meanwhile, to maximize the diversity of the dataset, an initial assessment of the similarity between scenarios, such as the data distribution sampling \cite{R27}, should be performed before the data collection, and the scenarios with more significant differences should be prioritized. {\color{black} In addition, given privacy and security considerations, federated learning \cite{FL_re1}, which employs model interaction rather than direct data mixing, represents a potential technical approach.}

In addition, considering the cost of the real data collection, efficient wireless data augmentation techniques, such as data conversion based on a priori physical properties of electromagnetic signals and data generation based on generative models, are worth exploring. Meanwhile, the significance of simulation data should be especially emphasized. Advanced simulation techniques, like ray-tracing, can generate high-fidelity wireless datasets at a low cost. Moreover, they can obtain additional information difficult to obtain in real measurements, such as the multi-path component of the channel and the unbiased user location. Therefore, the efficient mixing of simulated and measured data can significantly reduce the cost of the pre-training for wBAIM.

\subsection{Physics-Driven Learning Paradigm}\label{section5.2}

To build broad-domain generalization intelligence for BAIM, it is essential to establish a learning paradigm that enables the model to learn various regular phenomena. In wireless networks, the regulations of different tasks and scenarios are derived from the physical laws of electromagnetic waves. Therefore, the ideal learning paradigm for wBAIM should be physics-driven. {\color{black} On the one hand, incorporating known physical laws into the model's a priori design can significantly improve learning efficiency and generalization ability, reducing parameter redundancy and effectively controlling inference complexity. For instance, one could design the neural network's representation structure according to the physical structure of wireless data, define the optimization objective based on the distribution of wireless features, or select the activation function based on the mapping relationship of wireless information \cite{siren}.

On the other hand, the architecture design of wireless AI should shift from local module-oriented to global function-oriented, from task-driven to physics-driven. This transformation would facilitate better utilization of the physical correlations across multiple domains and dimensions of wireless data through necessary module reorganization. For example, in \cite{channel_deduction}, researchers proposed a novel channel acquisition method, channel deduction, by combining channel estimation and prediction. This approach overcomes the limitations of traditional channel acquisition models, leveraging the physical correlation of wireless data across multiple dimensions of space, time, and frequency, leading to a low-overhead, continuous autoregressive channel acquisition technique. }

\subsection{Wireless Scaling Laws}\label{section5.3}

Another typical feature of BAIM is that it continuously improves performance through the structure or scale expansion of the model, even producing the "emergence" effect. In language models, researchers have summarized the statistical relationships between parameter count, dataset size, computational cost, and loss value, broadly predicting the performance gains that can result from expanding model scale. This has also become an important theoretical foundation for the evolution of LLMs. This paper argues that increasing model size is also essential for the performance of wireless models, especially in terms of generalization, because sufficient neurons are necessary for the model to store diverse knowledge and perform complex reasoning. This principle is preliminarily confirmed in \cite{R28}.

Meanwhile, it is necessary to point out that the capability to jointly represent and process multi-modal, multi-user, and multi-scenario data is also crucial for wireless models to break through the current performance bottlenecks and generate the emergence effect of `more is different'. In wireless models, only limited real-time state information is available due to naturally constrained signaling costs, which differs greatly from language models, where sufficient information is usually accessible. This scarcity of known information can significantly restrict the model's performance, even if the model has extreme depth in information processing. Therefore, for wBAIM, the ability to simultaneously process multi-view information across a wide range of dimensions will significantly enhance their capacity to represent complex wireless environments, which will be a fundamental composition of the wireless scaling laws.

\subsection{Structural Prompt}\label{section5.4}

After learning a wealth of knowledge and intelligence, the model's efficient adaption to specific use cases is crucial for the deployment and application of BAIM. {\color{black} Common adaptation methods include fine-tuning, feature extraction, and prompting. Both fine-tuning and feature extraction require back-propagation and gradient descent in batches, which are demanding on hardware devices and computational resources. Meanwhile, this retraining process often demands considerable time, during which the model's intelligence expression may fail to meet current scenario requirements - a particularly critical issue for wireless communication systems with frequent environmental and configuration changes. Therefore, prompting, which does not require retraining, is the preferred adaptation method for wireless systems with highly decentralized computational resources.}

{\color{black} Unlike LLMs, where the prompt can linguistically be described in various forms, the ideal prompt for wBAIM should be strictly structured. The model must design specific data reading and processing patterns to extract essential features from the wireless prompt, facilitating an understanding of and adaptation to the target scenario. For instance, in wireless prompting, the data with different modalities should be processed through corresponding representation interfaces, and the orders and relationships between sub-elements must be strictly maintained during processing. This additional design is necessary due to the peculiarities of wireless data. As described in Section \ref{section4.1}, the meaning of wireless data depends on its strict logic and structure. In contrast, language has robust overall semantics and can convey core information even without strict logic and structure, which significantly simplifies the structure requirements for the prompt to LLMs. While in the development of wBAIM, researchers need to pay extra attention to developing and utilizing structural prompts.}

\subsection{Collaboration with AI-Agents}\label{section5.5}

\begin{figure*}[htbp]
    \centering
    \includegraphics[width=0.78\linewidth]{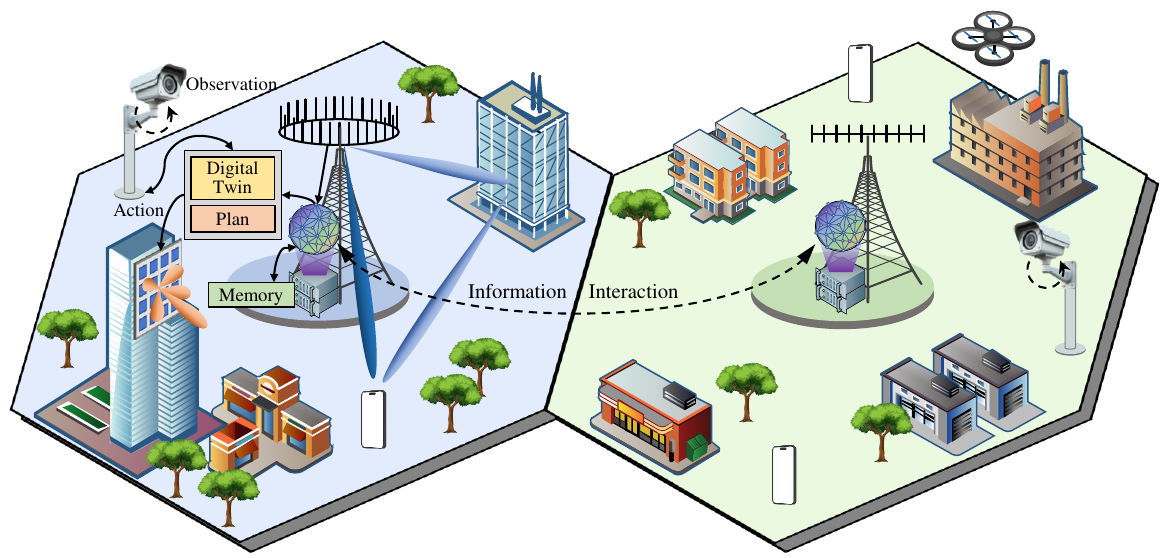}
    \vspace{0.3em}
    \caption{\color{black}An overall example of wBAIM's collaboration with AI-Agents.}
    \label{fig_agent}
\end{figure*}

After constructing wBAIM, the next step is to consider the integration with AI-agent, an important booster for expanding the practical applications of current BAIM technology. Building upon wBAIM’s deep understanding of wireless regulations, adding modules such as planning, observation, memorization, and small model tool libraries will create wireless intelligent agent, significantly enhancing the usability and flexibility of wireless intelligence. {\color{black} For instance, a planning module can improve wBAIM’s logical reasoning capabilities and optimize its global performance; an observation module can enable autonomous acquisition of necessary state information, reducing signaling overhead; a memory module can help the agent to evolve continuously and reduce repetitive thinking; and small-tool libraries can lower the energy consumption required for intelligent computations and optimize the system configurations. An overall example is shown in Fig. \ref{fig_agent}. Furthermore, the technical crux of coordinating such multifunctional modules lies in controlling the overall computational latency and achieving effective fusion and alignment of multi-modal information.}

{\color{black} Meanwhile, wireless intelligence is not an isolated intelligence. It can collaborate deeply with linguistic and visual intelligence to further facilitate the management of wireless systems. For example, by aligning wireless commands with natural language through techniques such as the projection layer, wireless agents can be effectively linked with other commonly used agents, improving the interactivity of wireless systems and reducing the technical barriers to operation.} In summary, the core objective of integrating wBAIM with AI agents is to avoid the passive and mechanical use of wireless intelligence, instead developing it into an intelligent wireless brain that can actively observe, control, optimize, and interact with system.

\section{Conclusion}\label{section6}
Ubiquitous intelligence will become an indispensable, fundamental enabling technology as the next-generation wireless networks evolve and improve their functionality, performance, and flexibility. To address the current challenges that wireless AI technologies face in terms of unit cost and reliability, BAIM, which has recently made significant breakthroughs in versatility and generalizability, has become a promising new approach. Therefore, the question of building a BAIM suitable for wireless networks has become a topic of great interest in wireless AI research.

In this study, we focus on creating a research roadmap for developing native wBAIM. We provide a detailed description of its scientific objectives, dissect the peculiarities, and indicate instructive methodologies for the evolution of wireless intelligence. This study can provide academic researchers and industry experts with insightful ideas on the evolution of wBAIM. With further efforts in these research directions, wBAIM is anticipated to create a revolution in the intelligence of entire wireless systems.
\\ \\
{\footnotesize\noindent
\textbf{Acknowledgement}~~This work was supported in part by National Natural Science Foundation of China under Grants 62394292 and 624B2129, Ministry of Industry and Information Technology under Grant TC220H07E, Zhejiang Provincial Key R\&D Program under Grant 2023C01021, and the Fundamental Research Funds for the Central Universities No. 226-2024-00069.}
\\



\end{document}